\documentclass[floatfix,aps,showpacs,amsmath,amssymb,twocolumn,eqsecnum,superscriptaddress]{revtex4-1}
\let\Twocolumn

\newif\ifTwocolumn
\ifx\Twocolumn\undefined
  \Twocolumnfalse
\else
  \Twocolumntrue
\fi
\ifTwocolumn
\fi
\usepackage{graphicx}
\usepackage{gensymb}
\usepackage{subfigure}
\usepackage{bm}
\usepackage{upgreek}
\usepackage{wasysym}
\usepackage{bibentry}
\usepackage[T1]{fontenc}
\usepackage{textcomp}
\usepackage{mathptmx}
\usepackage[scaled=0.92]{helvet}
\usepackage{courier}
\usepackage{setspace}
\usepackage{color}
\usepackage{epstopdf}
\usepackage{epsfig}
\usepackage{hyperref}
\usepackage{booktabs}
\usepackage{multirow}
\usepackage{amsmath}
\let\oldAA\AA
\renewcommand{\AA}{\text{\normalfont\oldAA}}
\definecolor{red_n}{rgb}{1.0, 0.0, 0.0}
\definecolor{brown_n}{rgb}{0.6, 0.4, 0.2}
\definecolor{cyan_n}{RGB}{0.0, 255.0, 255.0}
\definecolor{blue_n}{rgb}{0.0, 0.0, 1.0}
\definecolor{green_n}{rgb}{0.0, 0.5, 0.0}
\definecolor{orange_n}{RGB}{255.0, 127.0, 0.0}
\definecolor{magenta_n}{RGB}{255.0, 0.0, 255.0}
\definecolor{purple_n}{RGB}{128.0, 0.0, 128.0}
\definecolor{gray_n}{RGB}{128.0, 128.0, 128.0}
\definecolor{dark_green}{rgb}{0.0, 0.5, 0.0}
\newcommand{\eref}[1]{Eq.~\eqref{#1}}
\newcommand{\oncite}[1]{Ref.~[\onlinecite{#1}]}

\ifTwocolumn
  
\else
  
\fi

\begin{document}

\title{Behavior of the van der Waals force between a plate and a single-walled carbon nanotube under uniform hydrostatic pressure: a theoretical study}

\author{Galin~Valchev}
\email[Electronic address: ]{gvalchev@imbm.bas.bg}
\affiliation{Institute of Mechanics - Bulgarian Academy of Sciences, Academic Georgy Bonchev St. building 4, 1113 Sofia, Bulgaria}

\author{Peter~Djondjorov}
\email[Electronic address: ]{padjon@imbm.bas.bg}
\affiliation{Institute of Mechanics - Bulgarian Academy of Sciences, Academic Georgy Bonchev St. building 4, 1113 Sofia, Bulgaria}

\author{Vassil~Vassilev}
\email[Electronic address: ]{vasilvas@imbm.bas.bg}
\affiliation{Institute of Mechanics - Bulgarian Academy of Sciences, Academic Georgy Bonchev St. building 4, 1113 Sofia, Bulgaria}

\author{Daniel~Dantchev}
\email[Electronic address: ]{daniel@imbm.bas.bg}
\affiliation{Institute of Mechanics - Bulgarian Academy of Sciences, Academic Georgy Bonchev St. building 4, 1113 Sofia, Bulgaria}
\affiliation{Max-Planck-Institut f\"{u}r Intelligente Systeme, Heisenbergstrasse 3, D-70569 Stuttgart, Germany and
IV. Institut f\"{u}r Theoretische Physik, Universit\"{a}t Stuttgart, Pfaffenwaldring 57, D-70569 Stuttgart, Germany}

\date{\today}
%
%=============================================================================================
\begin{abstract}
We study the behaviour of the non-retarded van der Waals force between a planar substrate and a single-walled carbon nanotube, assuming that the system is immersed in a liquid medium which exerts hydrostatic pressure on the tube's surface, thereby altering its cross section profile. The shape of the latter is described as a continual structure characterized by its symmetry index $n$. Two principle mutual positions of the tube with respect to the substrate are studied: when one keeps constant the minimal separation between the surfaces of the interacting objects; when the distance from the tube's axis to the substrates bounding surface is fixed. Within these conditions, using the technique of the surface integration approach, we derive in integral form the expressions which give the dependance of the commented force on the applied pressure.
\end{abstract}
\maketitle
%
%=====================
\section{Introduction}\label{sec:Inrtoduction}
%=====================
%
Van der Waals forces (vdWf) are the dominant interactions, which govern the aggregation of electrically neutral atoms, molecules and complexes of such. At the basis of this type of forces are the dipole-dipole interactions, divided in : Keesom forces (i.e. between permanent dipoles) \cite{Keesom1921}; Debye forces (i.e. between permanent and induced dipoles) \cite{Debye1920}; and London forces (i.e. between instantaneously induced dipoles) \cite{FL1937}. As far as the theory explaining the origin of the first two is entirely based on the classical electrodynamics, the physical-mathematical apparatus used in the understanding of the manifestation of the latter is that of the quantum mechanics. In his original work London obtained an expression for the interatomic/intermolecular potential in fourth-order perturbation theory for the interaction of a dipole operator with a {\it fluctuating} electric field \cite{BKMM2009}. An important prerequisite for the emergence of the London-van der Waals interaction is the {\it correlation} between the {\it spontaneously arisen} and {\it induced} dipole moments in the particles considered. For distances greater than the so-called {\it retardation length} the correlation between the moments weakens, and the pair interaction falls even steeper with the distance, commonly known as Casimir-Polder interaction \cite{CP48}. A general theory of the non-retarded (London) and retarded (Casimir-Polder) forces, both known under the generic name {\it dispersion} interactions, was proposed by Lifshitz, Dzyaloshinskii and Pitaevskii in the case of plane parallel dielectric plates described by a frequency-dependent dielectric permittivity \cite{L56,DLP61}.

The piling research on the mechanisms and types of interactions between carbon structures is an indication of their degree of importance in the field of nanotechnology. In particular the dispersion forces are fundamentally important, when one examines the interactions between pair of carbon structures, including graphene \cite{BMP2018,NoIn2018,LCWXLDZ2018}, fullerenes \cite{HGZM2016,CYPT2016,COP2017} and carbon nanotubes (CNTs) \cite{ZPC2010,SAP2018,MHH2018}. When it comes down to discussing the stability \cite{VoZh2010,ZJWR2015}, vibration modes \cite{WSLSG2012,KhHa2012} and mechanical performance \cite{LYZ2017} of CNTs, the understanding of vdW interactions becomes essential.

In the past two decades, the subject on CNTs deformation under different mechanical loads (e.g., external hydrostatic pressure) has been under considerable interests.
In a series of papers, Ou-Yang and co-authors \cite{Ou-Yang1997,Tu2002,Tu2008} have proposed and developed a model describing the equilibrium shape of CNTs as the continuum limit of the lattice model proposed by Lenosky {\it et al.} \cite{Lenosky1992}. In particular, the main finding of their work is that the expression for the curvature elastic energy of a CNT is the same as that of fluid membranes \cite{Hel1973,DH1976} and solid shells \cite{LL2012}. The evolution of the cross-section profile of a single-walled CNT (SWCNT) subject to uniform hydrostatic pressure was studied by carrying out molecular dynamics simulations \cite{Zang2004,Tangney2005,ZAL2007} based on different inter-atomic potentials, density-functional theory calculations \cite{CYGL2003}, as well as solving numerically the shape equation \cite{Xie1996} derived within the aforementioned continuum mechanics model \cite{OYH1987}. It is noteworthy that the results obtained from these approaches are in an excellent agreement. Later on, all solutions of the foregoing shape equation determining cylindrical equilibrium configurations of SWCNT under uniform hydrostatic pressure were found and given, together with the expressions for the corresponding position vectors, in explicit analytical form \cite{Vassilev2008,Djondjorov2011,Mladenov2013,Vassilev2015}.

When the geometry of a CNT changes it affects not only its intrinsic properties but also the interactions with the surrounding objects. This, in turn, affects the performance of nano-devices composed out of CNTs that might operate under extreme conditions. Therefore, it is crucial to gain knowledge on the dependance of the vdWf between radially non-circular CNTs and objects of various geometries in terms of distances, relative orientations, interaction potential etc.

The aim of the current article is to describe the dependance of the vdWf between a CNT and a planar substrate under the conditions when applied external mechanical load alters above certain magnitude the stress-free cross section of the CNT. To do so, we base our study on results already reported in Refs. \cite{Djondjorov2011} and \cite{DV2012}, as the first concerns the description of the radial cross-section change of a CNT under uniform hydrostatic pressure (see Subsec. \ref{sec:CrossSec}), and the second provides the means to calculate the vdWf between the deformed tube and a planar substrate (plate), say stack of flat graphene sheets (see Subsec. \ref{sec:SIAapp}). Using this knowledge, in Sec. \ref{sec:SWCNTforce} we provide in integral form the expressions for the tube-plate force per unit length. The observed behaviour of this force as a function of the applied pressure, after the numerical evaluation of these expressions, is analysed in details in Sec. \ref{sec:ResAndDisc}. We conclude the exposition with a summary and discussion section -- Sec. \ref{sec:Conclusions}.

%-----------------------------
\begin{figure}[h]
\centering
\includegraphics[width=\columnwidth]{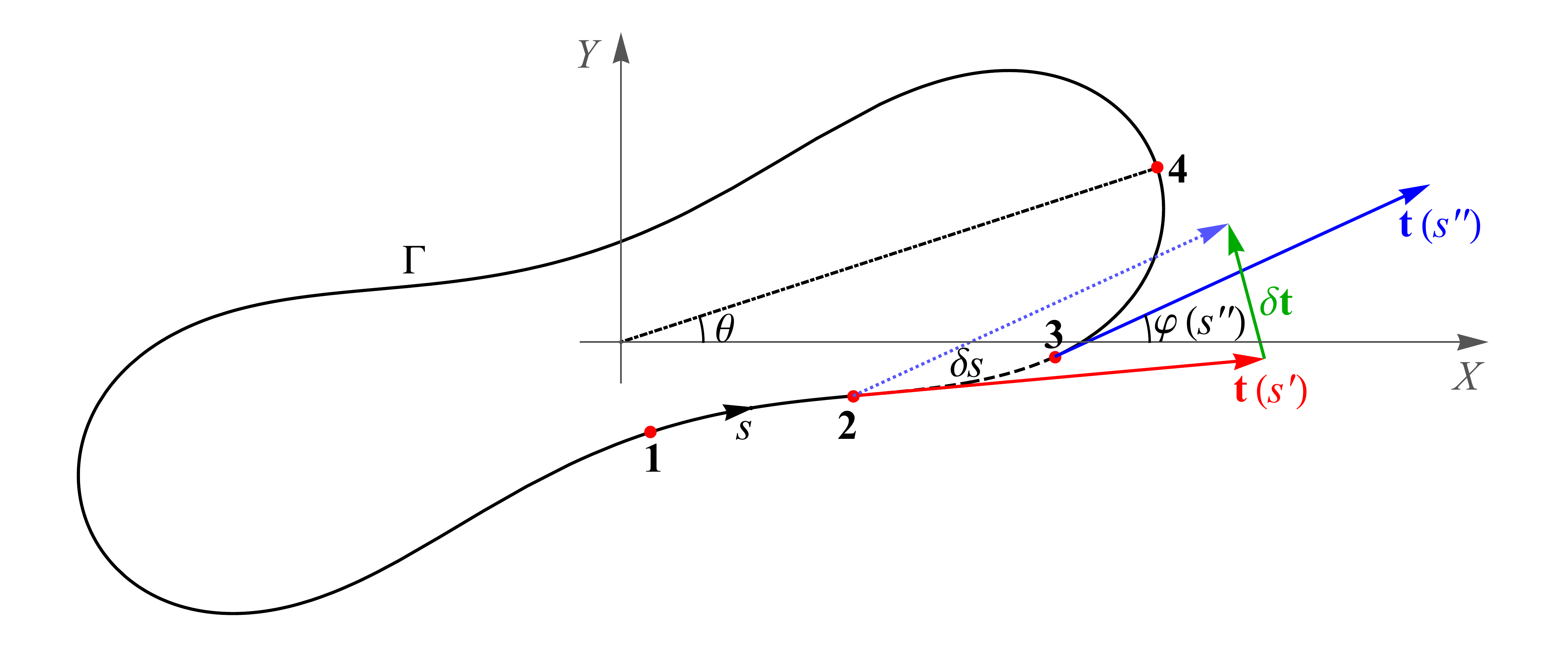}
  \caption{Schematic depiction of the continual non-radial cross section profile $\Gamma$ of a CNT under uniform mechanical load. Here $\theta$ is the inclination angle with respect to the $X$-axis (the plane bounding the substrate surface), $s$ is the arc length measured in direction from point $\rm{\bf{1}}$ - $x(0),y(0)$ to point $\rm{\bf{4}}$ - $x(nT/4),y(nT/4)$, and the curvature at any point, say point $\rm{\bf{2}}$, is defined as $\kappa(s')=\lim_{\delta s\rightarrow 0}|\delta{\bf{t}}/\delta s|$, where $\delta{\bf{t}}={\bf{t}}(s'+\delta s)-{\bf{t}}(s')$, $\delta s=s''-s'$, with $s''$ being the arc length value at point $\rm{\bf{3}}$. Here the unit tangential vector is given by ${\bf{t}}(s)=\cos\varphi(s){\mathbf{e}}_{X}+\sin\varphi(s){\mathbf{e}}_{Y}.$ }
  \label{fig:cross_section}
\end{figure}
%-----------------------------
%--------------------------------------------------------------------------------------------------------------------------------------------------------------------------
\begin{figure*}[th!]
  \centering
\mbox{\subfigure{\includegraphics[width=7.0 cm]{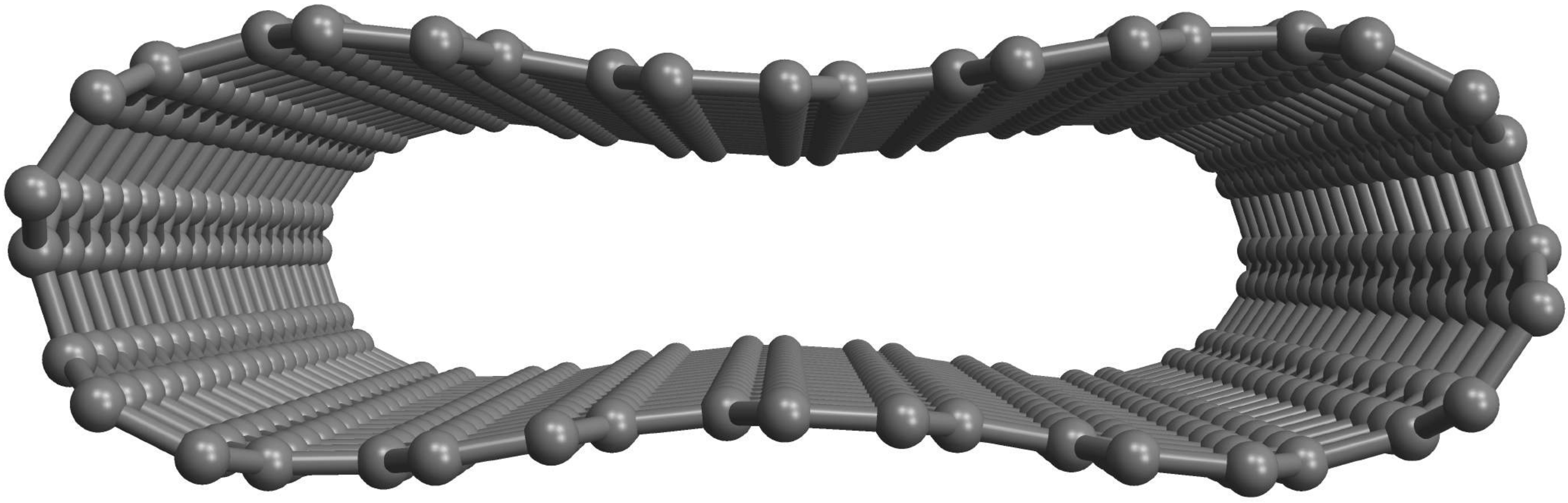}}\quad
      \subfigure{\includegraphics[width=5.2 cm]{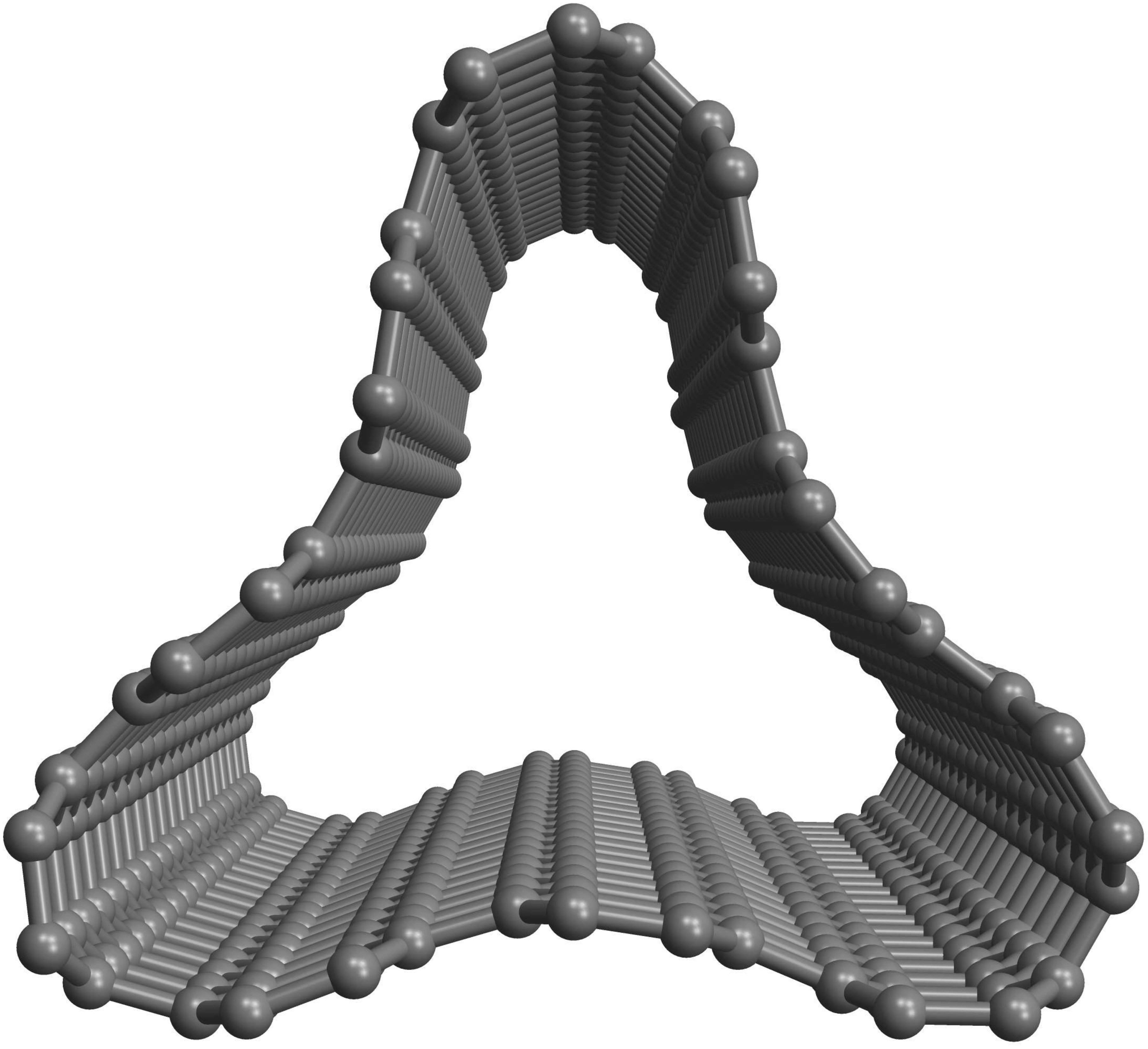}}\quad
      \subfigure{\includegraphics[width=4.2 cm]{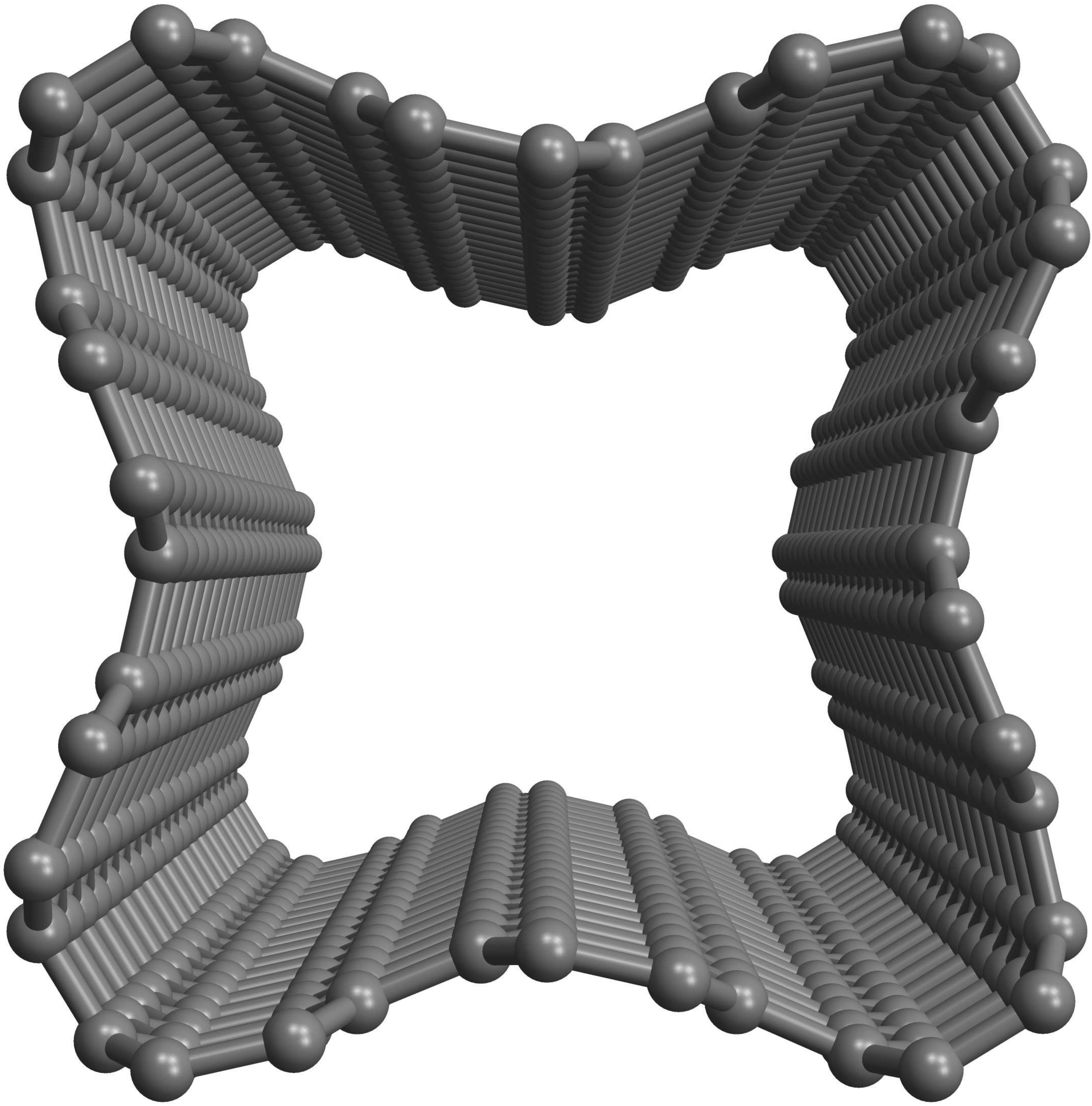}}}
  \caption{Cross section view of armchair $(10,10)$ SWCNTs (each of length 20 unit cells) at different values of the pressure $\sigma$: (left) $4.0$; (middle) $11.0$; (right) $20.0$. The coordinates of the carbon atoms in the initial (non-deformed/ring) configuration are gained using the carbon nanotube generation applet, written by Veiga {\it et al.}  \cite{VTF2008}. The visualized shapes are then obtained implementing the discussed theoretical model in \textsc{Mathematica}$^\circledR$. The carbon-carbon bond length in the initial configuration was chosen $1.42\ \AA$ \cite{BZRLM2005}.}
  \label{fig:cross_section_CNT}
\end{figure*}
%--------------------------------------------------------------------------------------------------------------------------------------------------------------------------
%
%-----------------------------
\begin{figure}[t!]
\centering
\includegraphics[width=\columnwidth]{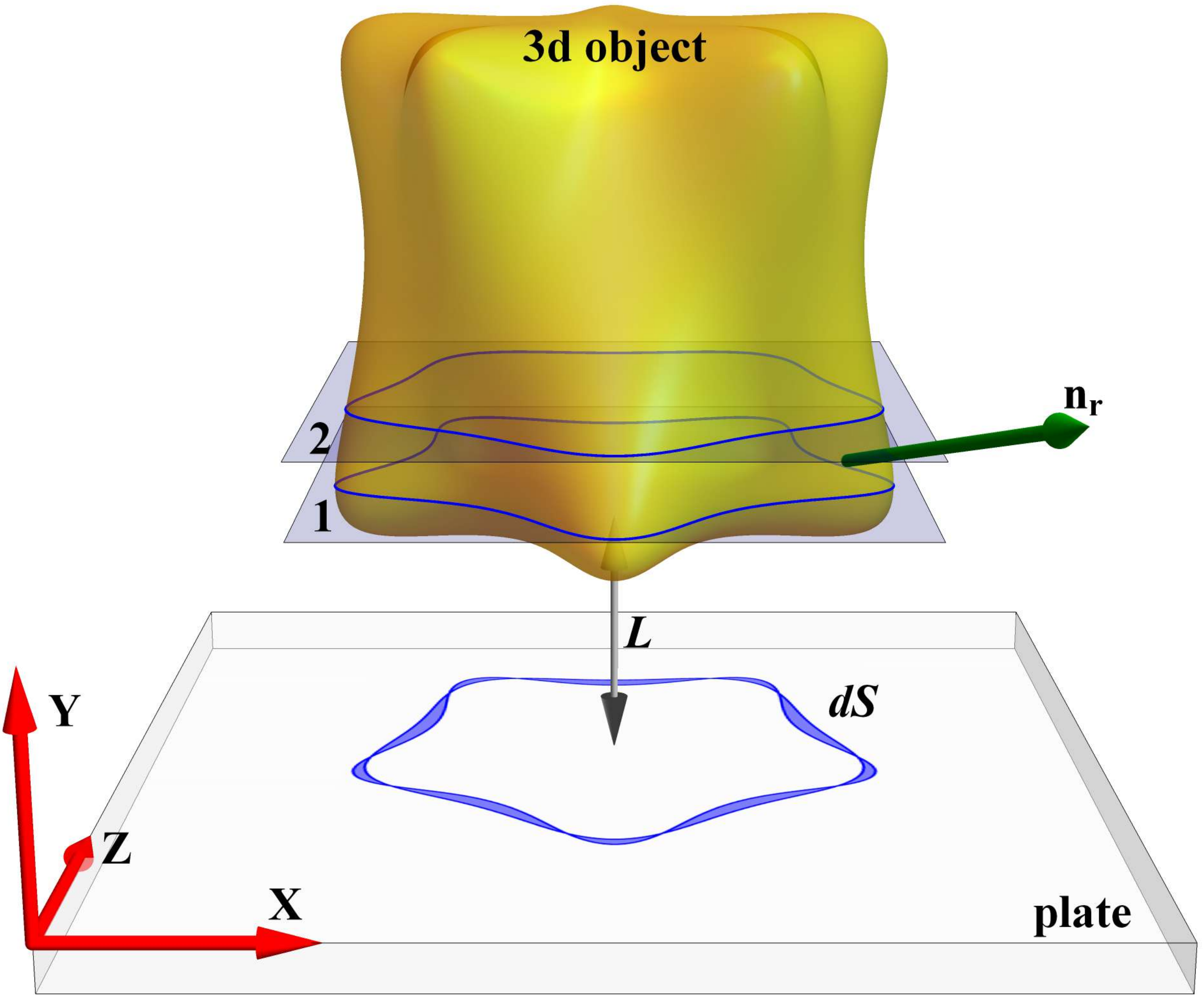}
  \caption{Explanatory depiction of the SIA applied for the interaction between arbitrary in shape 3d object and a flat (half-infinite/thick) plate, separated by a gap of length $L$. The infinitesimal projected area element $\mathrm{d}S$ is calculated by simply subtracting the space $S_{g_{2}}$ enclosed by contour $\mathbf{2}$, whose general equation can be written as $g_{2}(x,y+\mathrm{d}y,z)=0$, and that, $S_{g_{1}}$, enclosed by contour $\mathbf{1}$ - $g_{1}(x,y,z)=0$, for some fixed value of $y$. Here we note that when $S_{g_{2}}>S_{g_{1}}$, the $y$-component of the resultant unit normal vector $\mathrm{\mathbf{n_{r}}}$ to the infinitesimal strip formed between $g_{1}$ and $g_{2}$, is negative, which defines a surface region of the particle that "faces towards" the plate - $A_S^{\rm to}$, and $|\mathrm{\mathbf{n_{r}}}|\mathrm{d}S>0$. For similar reasons for $S_{g_{2}}<S_{g_{1}}$, one has a surface region that "faces away" from the plate - $A_S^{\rm away}$, and $|\mathrm{\mathbf{n_{r}}}|\mathrm{d}S<0$.}
  \label{fig:SIA3d}
\end{figure}
%-----------------------------
%--------------------------------------------------------------------------------------------------------------------------------------------------------------------------
\begin{figure*}[t!]
  \centering
\mbox{\subfigure{\includegraphics[height=5.55 cm]{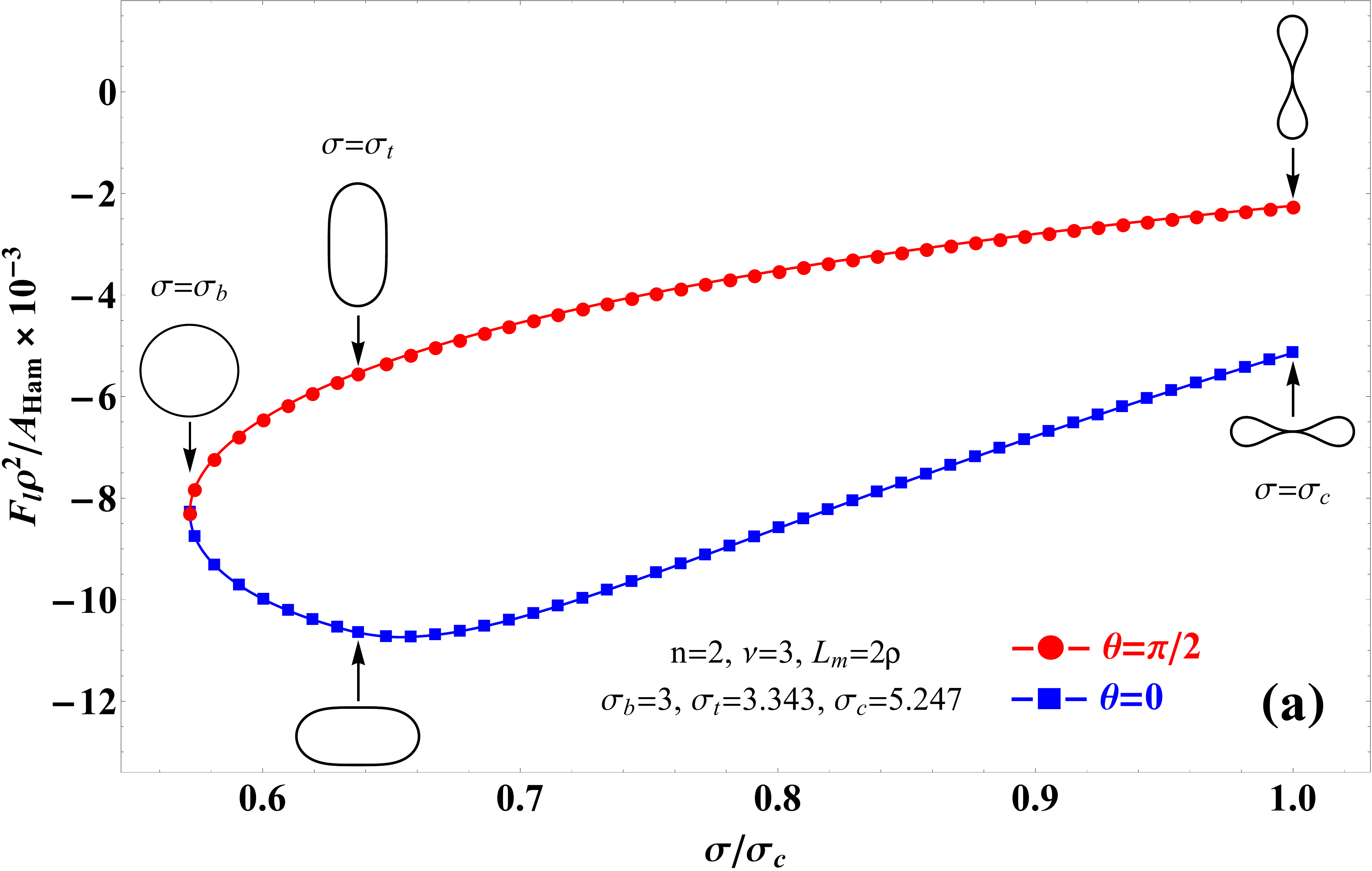}}\quad
      \subfigure{\includegraphics[height=5.55 cm]{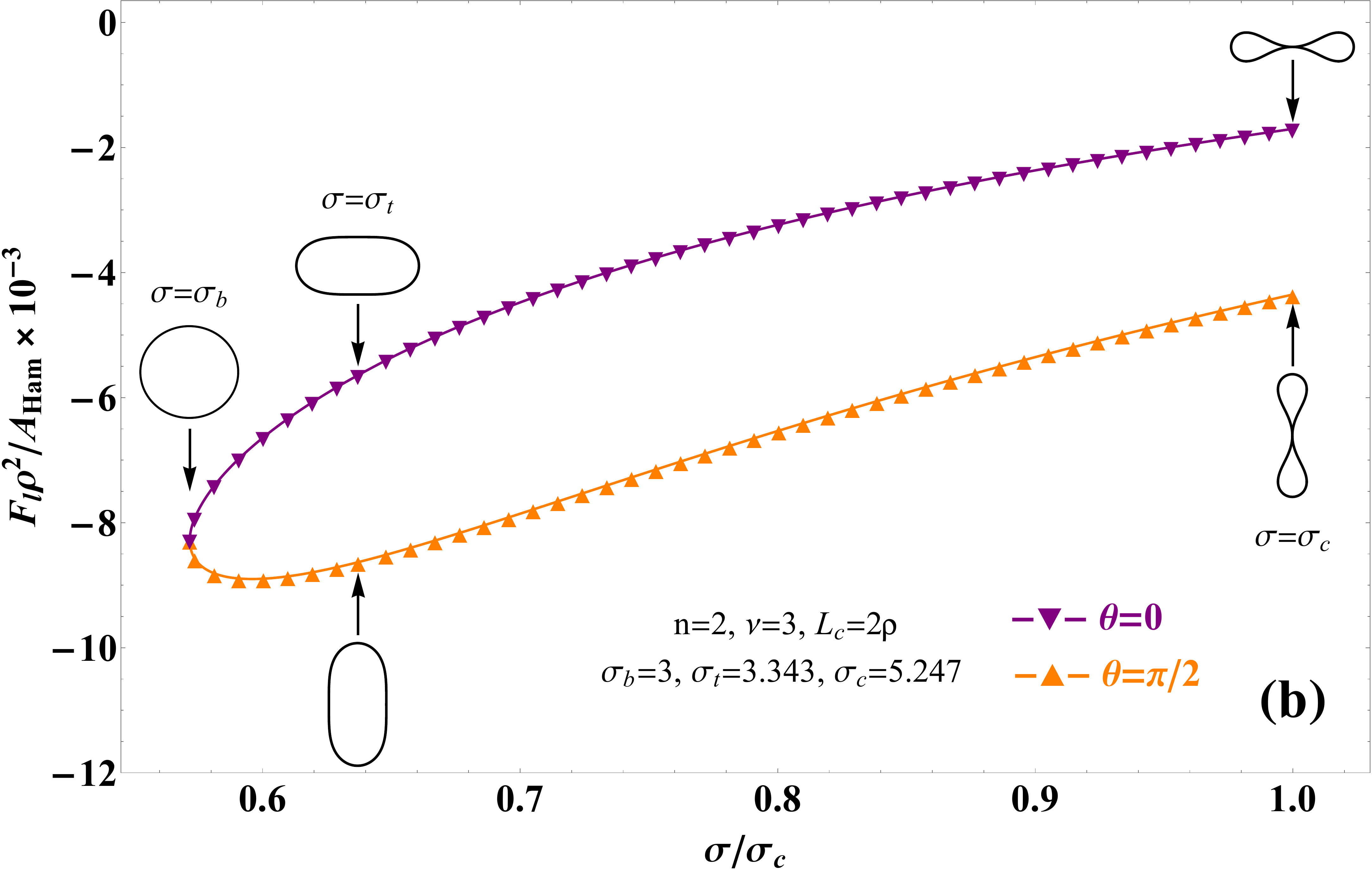}}}\\
\mbox{\subfigure{\includegraphics[height=5.55 cm]{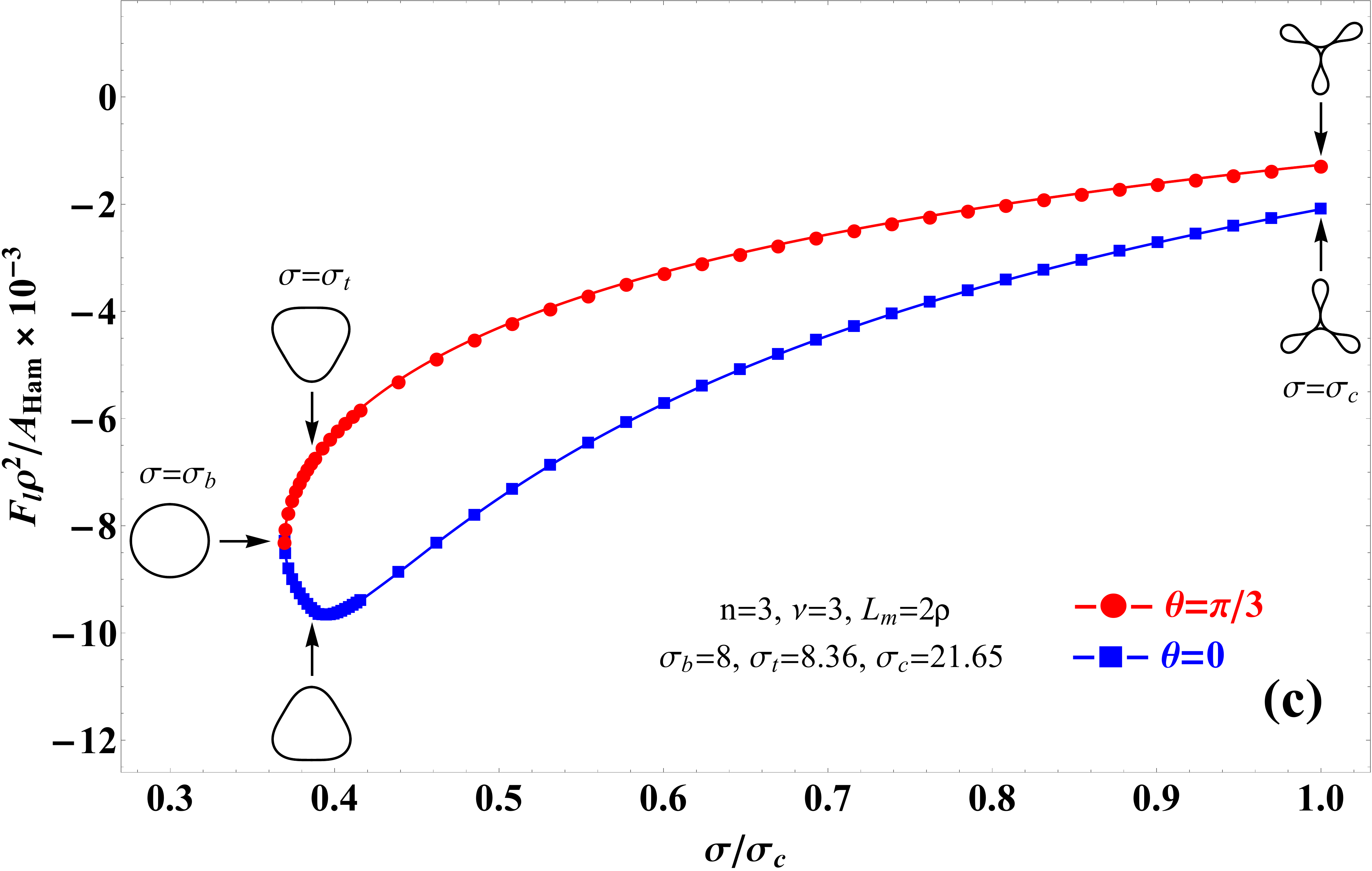}}\quad
      \subfigure{\includegraphics[height=5.55 cm]{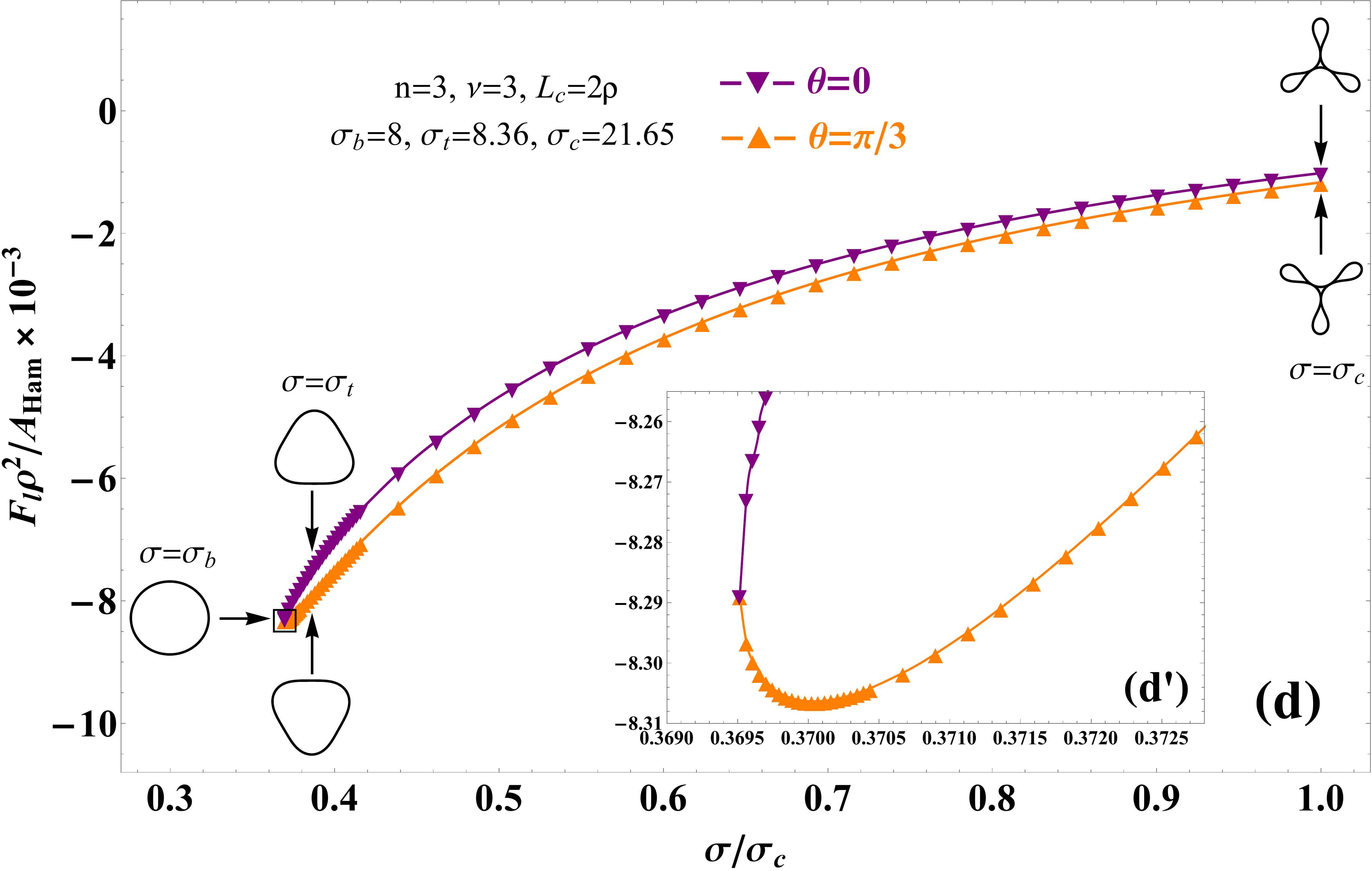}}}\\
\mbox{\subfigure{\includegraphics[height=5.55 cm]{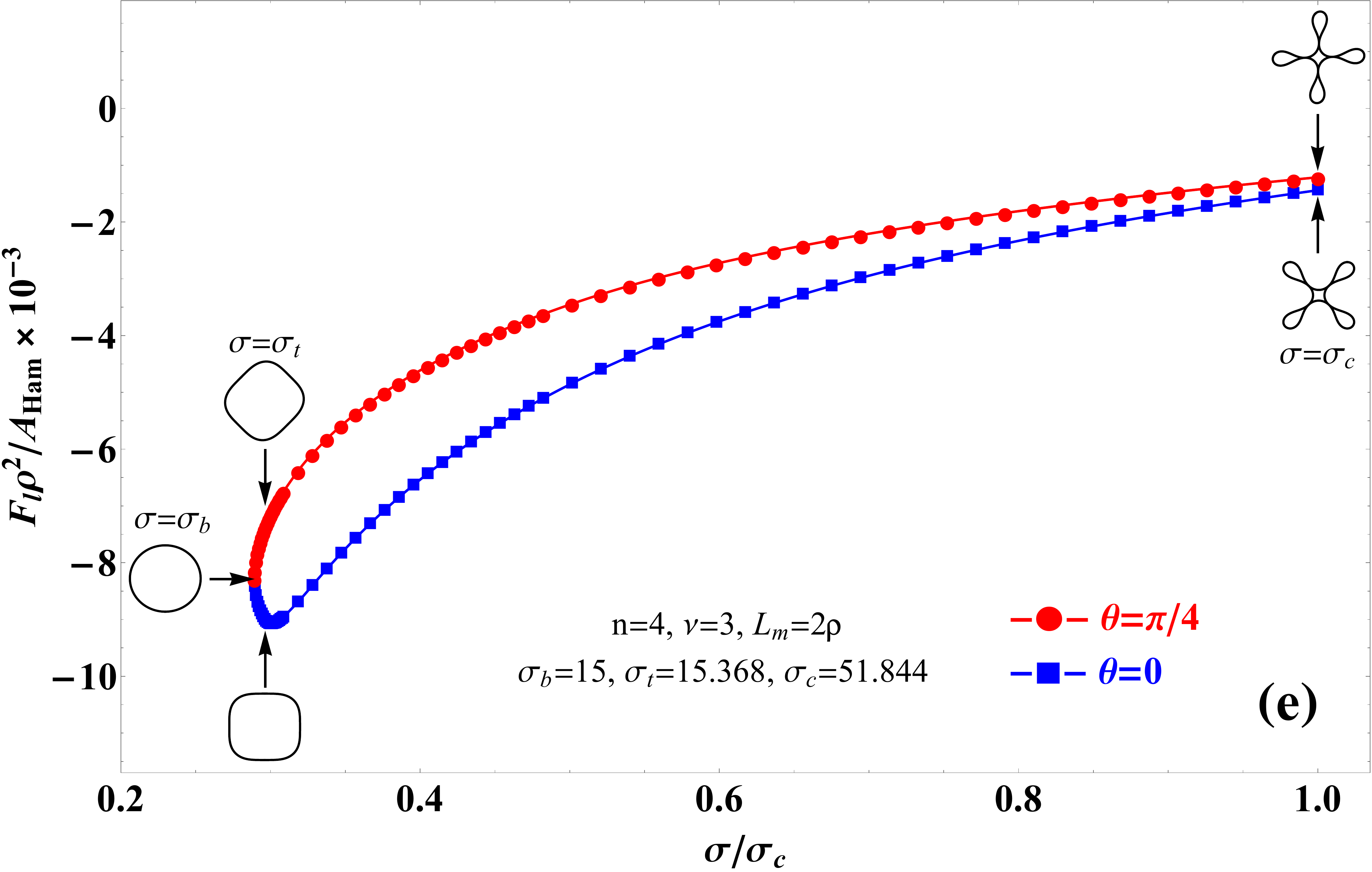}}\quad
      \subfigure{\includegraphics[height=5.55 cm]{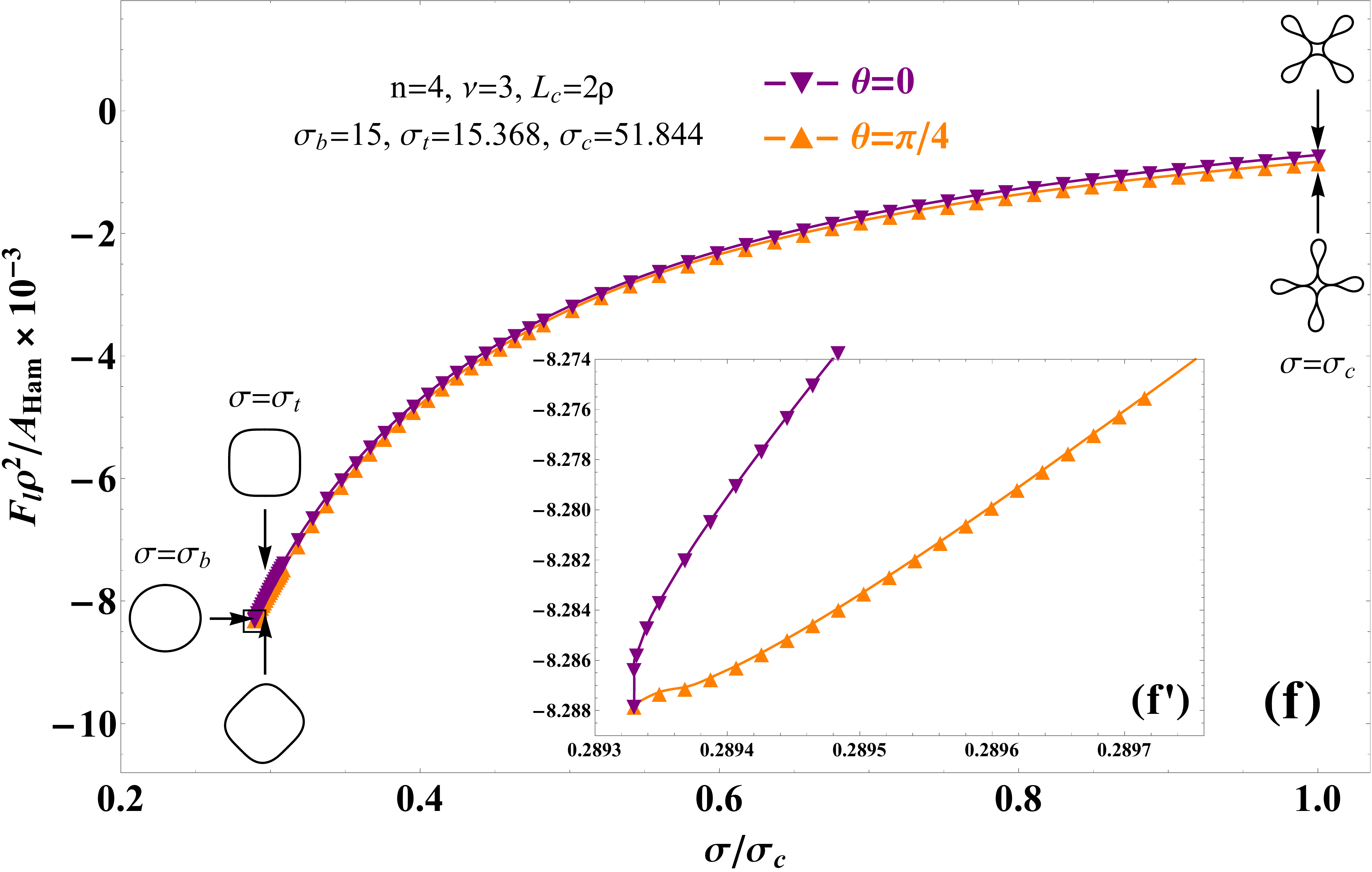}}}
  \caption{Dependance of the dimensionless standard $(\nu=3)$ van der Waals tube-plate force $\Phi_{l}\equiv F_{l}\rho^{2}/A_{\rm Ham}$ [Eqs. (\ref{sbstn234}), (\ref{stscn234})-(\ref{stscn4})] from the dimensionless pressure $\sigma$ relative to $\sigma_{c}$ at any realized transverse contour of $n$-fold symmetry. For the cases presented, the tube-plate separation is considered fixed as $L_{m}/\rho=L_{c}/\rho=2$. We choose to depict the force-pressure dependance only in the limiting orientations $\theta=0$ and $\theta=\pi/n$ of the tube with respect to the substrate's surface, since at any other $\theta$, the value of the force lies between these two points for any fixed $\sigma/\sigma_{c}$. The corresponding cross section shapes occurring at the commented in the text key values of $\sigma$ for any considered $n$ are also illustrated.}
  \label{fig:total_forces}
\end{figure*}
%--------------------------------------------------------------------------------------------------------------------------------------------------------------------------
%=====================
\section{Theoretical background}\label{sec:TheorBack}
%=====================
%
%=====================
\subsection{Analytical description of the equilibrium cross section}\label{sec:CrossSec}
%=====================
%
Within the model discussed in \cite{Djondjorov2011}, when the pressure $p$ is constant in magnitude and acts as an external uniformly distributed force along the inward normal vector to the surface of a carbon nanotube, the Cartesian coordinates of the tube's cross section $\Gamma$, parameterized by the arc length $s$, are given by
\begin{equation}
\begin{bmatrix}
x(s) \\
y(s)
\end{bmatrix}
=
\varsigma^{-1} \begin{bmatrix}
\cos\varphi(s) & \sin\varphi(s)\\
\sin\varphi(s) & -\cos\varphi(s)
\end{bmatrix}
\begin{bmatrix}
{\rm d}\kappa(s)/{\rm d}s \\
\frac{1}{2}\kappa^{2}(s)-\mu
\end{bmatrix}.
\label{initeq}
\end{equation}
Here $\varsigma=p/D$ is the pressure (in units ${\rm m^{-3}}$), with $D$ being the bending (flexural) rigidity (in units ${\rm Pa\cdot m^{3}}$) of the cross section ring and $\mu=(\alpha^{2}+\beta^{2}+\gamma^{2}+\alpha\beta+\beta\gamma+\alpha\gamma)/4$. The expression for the slope angle $\varphi(s)$ is as follows
\begin{eqnarray}\label{slopeangle}
\varphi(s)&&=\frac{A\beta-B\alpha}{A-B}s+\frac{\alpha-\beta}{2\lambda\sqrt{k^{2}+C}}\arctan\left[\sqrt{k^{2}+C}\frac{{\rm sn}(\lambda s,k)}{{\rm dn}(\lambda s,k)}\right]\nonumber\\&&+\frac{(A+B)(\alpha-\beta)}{2\lambda(A-B)}\Pi\left[-C,{\rm am}(\lambda s,k),k\right],
\end{eqnarray}
where $C=(A-B)^{2}/4AB$, $\lambda=\sqrt{AB}/4$, ${\rm sn}(\cdot,\cdot)$, ${\rm dn}(\cdot,\cdot)$, ${\rm am}(\cdot,\cdot)$ and $\Pi(\cdot,\cdot,\cdot)$ denote the elliptic sine, delta amplitude, Jacobi amplitude and the incomplete elliptic integral of the third kind, respectively. In \eref{slopeangle} the elliptic modulus $k$ is given by
\begin{equation}\label{elmod}
k=\sqrt{\frac{1}{2}-\frac{4\eta^{2}+(3\alpha+\beta)(\alpha+3\beta)}{2AB}},
\end{equation}
with $A=\sqrt{4\eta^{2}+(3\alpha+\beta)^{2}}$ and $B=\sqrt{4\eta^{2}+(\alpha+3\beta)^{2}}$. For the curvature $\kappa(s)$ of $\Gamma$ the model shows that
\begin{equation}\label{curvature}
\kappa(s)=\frac{(A\beta+B\alpha)-(A\beta-B\alpha){\rm cn}(\lambda s,k)}{(A+B)-(A-B){\rm cn}(\lambda s,k)},
\end{equation}
where ${\rm cn}(\cdot,\cdot)$ is the elliptic cosine. The constants $\alpha,\ \beta,\ \gamma$ and $\delta$, appearing in some of the above expressions are the roots of the polynomial $P(\kappa)=-\frac{1}{4}\kappa^{4}+\mu\kappa^{2}+2\varsigma\kappa+\varepsilon$ and are explicitly given by
\begin{subequations}\label{abgd}
\begin{equation}\label{ab}
\alpha=\frac{4\varsigma}{\eta^{2}+q^{2}}-q,\ \beta=2q+\alpha,
\end{equation}
\begin{equation}\label{gd}
\gamma=-\frac{\alpha+\beta}{2}+{\rm i}\eta,\ \delta=\overline{\gamma},
\end{equation}
\end{subequations}
where $\eta$ and $q$ are positive real numbers and the bar over $\gamma$ designates complex conjugation. The free term in $P(\kappa)$ is equal to: $\varepsilon=-2\varsigma\kappa^{\circ}-\varsigma^{2}c-\mu^{2}$, where $\kappa^{\circ}=1/\rho$ is the curvature of the stress-free cross section, which is supposed to be a circle of radius $\rho$, and $c$ is an arbitrary constant. The inspection of Eqs. (\ref{initeq})-(\ref{abgd}), shows that in order to obtain the coordinates of the carbon nanotube cross section at certain value of $\varsigma$, i.e., pressure, one needs to determine the parameters $\eta$ and $q$. The system of equations which solution determines the values of $\eta$ and $q$ is the following: $\varphi(T)=\pi/2n$ and $T=2\pi\rho/n$, with $T=4\lambda^{-1}{\rm K}(k)$, where ${\rm K}(\cdot)$ denotes the complete elliptic integral of the first kind. The first equation represents the closure condition of $\Gamma$, while the second takes into account that the length $C$ of the cross section is fixed and does not change upon deformation. All the characteristics of the cross section contour commented so far are visualized on Fig. \ref{fig:cross_section}.

Since $\varsigma$ is the main quantity which determines the cross section profile (see Fig. \ref{fig:cross_section_CNT}), one can pinpoint several key values of it at which essential changes of that shape occurs. Introducing the dimensionless pressure $\sigma\equiv\rho^{3}\varsigma$, the model predicts that for $\sigma\leq\sigma_{b}\equiv n^{2}-1$ the tube's cross section is a circle of radius $\rho$, with the curvature at any point positive and inversely proportional to $\rho$. Here $\sigma_{b}$ is the so-called "\textit{buckling} pressure" and $n\geq2$ is an integer which is interpreted as the number of symmetry axes, of a non-circular cross section, with respect to which the shape in question is symmetric upon reflection. When the value $\sigma_{b}$ is exceeded the cross section is no longer only a circle, and its curvature now depends on the point at which it is being measured. The first value of $\sigma$ at which a given cross section of $n$-fold symmetry has exactly $n$ points at which the curvature is zero, i.e. when $\alpha=0$, will be termed "\textit{threshold} pressure", and designated by $\sigma_{t}$. Within the theory commented so-far $\sigma_{t}=q(\eta^{2}+q^{2})/4$. Further increase of $\sigma$ results, in purely mathematical sense, to contact between opposite points on a CNT's contour. This value is designated by $\sigma_{c}$ and dubbed "\textit{contact} pressure". Its amount for various $n$ is tabulated in \cite{Djondjorov2011} (see Table 1 there).

%=====================
\subsection{General idea about the SIA approximation}\label{sec:SIAapp}
%=====================
Here we briefly remind the technique of "the surface integration approach" (SIA), introduced in Ref. \cite{DV2012}, which will be used in calculating the force between a planar substrate and a single-walled carbon nanotube.

In 1934 the soviet scientist B. Derjaguin was the first to propose an approach \cite{D34} for calculating geometry dependent interactions in systems where at least one of the objects has a non-planar geometry. Depending on the research field in which this technique is used, it is known as Derjaguin approximation (DA) in colloidal science and proximity force approximation in studies of the QED Casimir effect (see p. 79 in \oncite{ButtKap2018}). In particular the DA focuses on relating the interaction force/potential between two gently curved colloidal particles with the knowledge for that between a pair of parallel plates $f^{\parallel}_{{\cal A}}$. An important feature of this technique is that it is \textit{only} applicable if the separation distance between the interacting objects is {\it much smaller} than their geometrical characteristics.

In order to overcome this inconvenient condition the co-called "surface integration approach" (SIA), reported in Ref. \cite{DV2012}, was developed. The main advantage of this new approach over the DA is that one is no longer bound by the restriction that the interacting objects must be much closer to each other than their characteristic sizes. Here we must also note that both DA and SIA are strictly valid if the interactions involved can be described by pair potentials, i.e. are additive. Even though the London-van der Waals forces, do not count as such \cite{VWR2016}, their non-additive behaviour is accounted {\it{only}} by the Hamaker constant (see p. 58 in \oncite{ButtKap2018}), and hence one can make use of the DA and/or SIA to evaluate these forces between objects of various geometries. Within the SIA the interaction force $\mathbf{F}^{B,|}$ between an object (say a colloid particle) $B$ of arbitrary shape and a flat surface bounded by the $(x,z)-$plane of a Cartesian coordinate system, is determined by subtracting from the contributions stemming from the surface regions $A_S^{\rm to}$ of the particle that "face towards" the plane those from the regions $A_S^{\rm away}$ that "face away" from it (see Fig. \ref{fig:SIA3d}). Here $A_S^{\rm to}$ and $A_S^{\rm away}$  are the projections of the corresponding parts of the surface of the body on the $(x,z)-$plane. Due to the symmetry of the mutual orientation of the interacting objects, the $x$ and $z$ components of $\mathbf{F}^{B,|}$ are zero and hence
\begin{equation}\label{SIAgeneralsimple}
\mathbf{F}^{B,|} =\oint_{S}f^{\parallel}_{{\cal A}}\mathbf{n_{r}}\mathrm{d}S.
\end{equation}
Note that the expression Eq. (\ref{SIAgeneralsimple}) takes into account that the force on a given point of $S$ is {\it along the normal} to the surface at that point (for details see Section 2 in Ref. \cite{DV2012}). It is clear that if one takes into account only the contributions over $A_S^{\rm to}$ the results will be an expression very similar to one obtained using the DA.

In the next section we give the explicit form of \eref{SIAgeneralsimple} when the object $B$ is a CNT whose cross-section symmetry index $n=2,3 \ \text{and}\ 4$.

%=====================
\section{The force per unit length between a SWCNT of non-circular cross section and a thick planar substrate within the SIA approximation}\label{sec:SWCNTforce}
%=====================
%-----------------------------
\begin{figure}[h]
\centering
\includegraphics[width=\columnwidth]{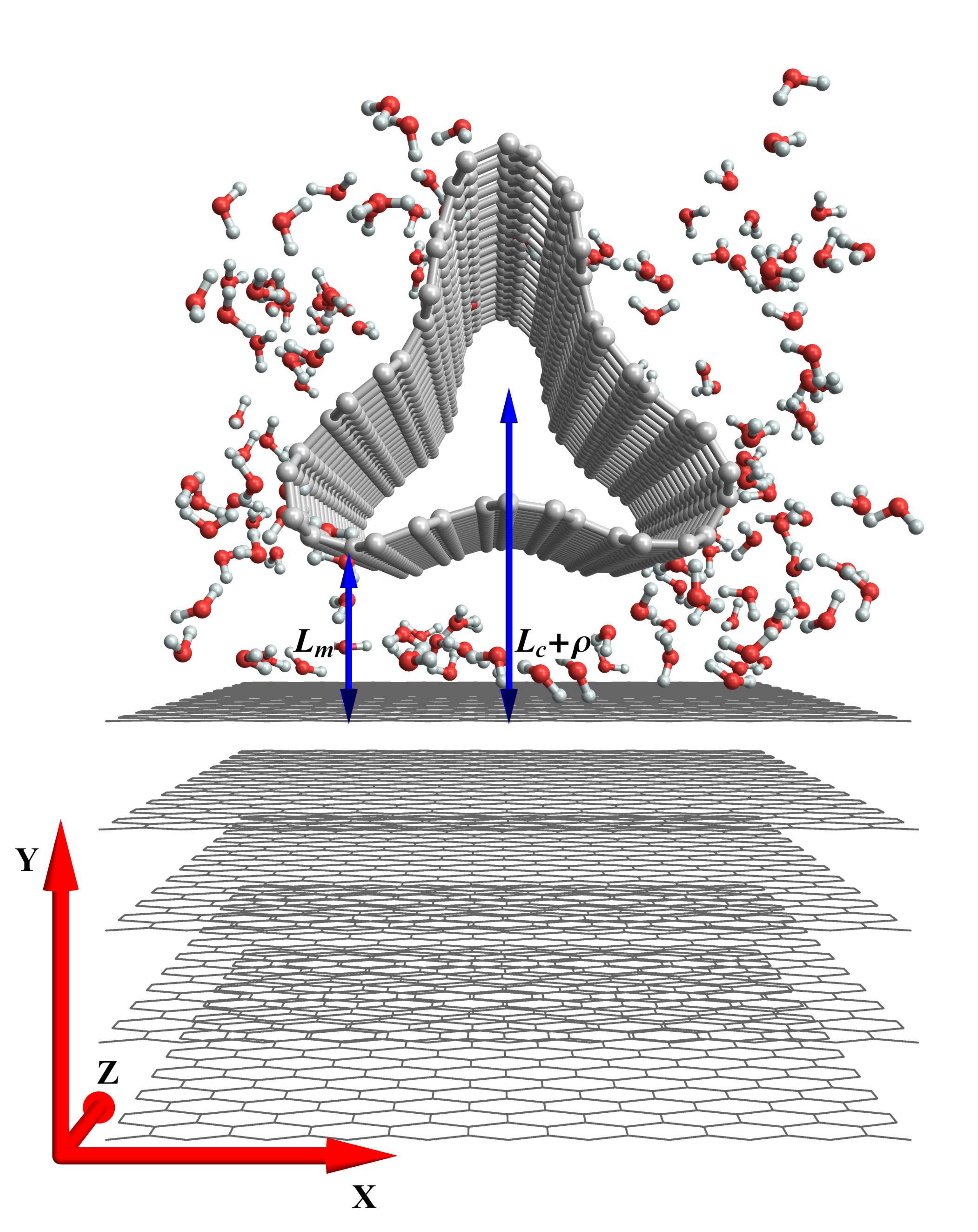}
  \caption{3D illustration of the considered mutual orientation between a CNT and a substrate. In this depiction the latter is sketched as a stack of flat graphene sheets, apart from one another by $\sim 3.2\ \AA$ \cite{PWB2008}. The liquid medium (shown here as separate water molecules) in which the tube is considered immersed in, exerts an external uniformly distributed force along the inward normal vector to the tube's surface, which changes the geometry of the latter. The distances $L_{m}$ and $L_{c}$ are the two principal ones with respect to which the interaction between the CNT and the substrate is being studied, as $L_{m}$ designates the minimal gap between the surfaces of the objects involved and $L_{c}+\rho$ is the separation between the substrate's surface and the center of the tube.}
  \label{fig:illustration}
\end{figure}
%-----------------------------
Let us now consider a SWCNT of length $l$ which transverse contour can be described, in terms of the theory commented in Sec. \ref{sec:CrossSec}, as a shape
of $n$-fold symmetry, realized for some value of $\sigma$ between $\sigma_{b}$ and $\sigma_{t}$. Then, the force per unit length $F_{l}\equiv F/l$ between the carbon nanotube and a planar substrate, within the assumptions of the SIA is given by
  \begin{equation}\label{sbstn234}
      F_{l}(L)=2\int_{s_{w}}^{(nT/2)+s_{w}}\frac{\mathrm{d}x_{\theta}(s)}{\mathrm{d}s}f^{\parallel}_{{\cal A}}(s_{w},\theta|L)\mathrm{d}s,
      \end{equation}
with $s_{w}=0$ for a non-rotated tube, i.e. $\theta=0$, and $s_{w}=T/2$ when one studies the case of a tube rotated clockwise to half its symmetry angle, i.e. $\theta=\pi/n$, around the central axis. In \eref{sbstn234} the appearing force per unit area is given by
\begin{equation}\label{FpuA}
  f^{\parallel}_{{\cal A}}(s_{w},\theta|L)=-\frac{2A_{\rm Ham}}{\nu(\nu+1)\pi}\frac{\xi_{\rm ret}^{\nu-3}}{\Delta_{\theta}^{\nu}(s)},
\end{equation}
where $\Delta_{\theta}(s)\equiv y_{\theta}(s)+\Lambda(s_{w})+L$ is the distance between an elementary projected area element ${\rm{d}}S$, characterized by its value $s$, from the CNT's exterior and the surface of the plate. The separation $L$ is such that if one considers the case of fixed minimal distance between the surface of the tube and that of the plate, $L\equiv L_{m}$ and $\Lambda(s_{w})=-y_{\theta}(s_{w})$, whereas when one is interested in the interaction at fixed tube center-plate separation, $L\equiv L_{c}$ and $\Lambda(s_{w})=\rho$ (see Fig. \ref{fig:illustration}). The constants $A_{\rm Ham}$ and $\xi_{\rm ret}$ which appear in \eref{FpuA} are the so-called Hamaker constant and retadration length, respectively. The first depends only on the material characteristics of the interacting objects and the medium they are immersed in, but do not depend on any geometrical characteristics in the system, whereas the second constant is a medium specific quantity, which is a measurement for the distance at which the retardation effects are felt. The exponent $\nu$ is a characteristic for the decay of the interaction, as $\nu=3$ corresponds to the standard (London) van der Waals interaction, while $\nu=4$ describes the retarded (Casimir) one. The relation between the rotated and non-rotated coordinates is as follows
\begin{equation}\label{rotmat}
     \begin{bmatrix}
         x_{\theta}(s) \\
         y_{\theta}(s)
        \end{bmatrix}
        =
        \begin{bmatrix}
         \cos\theta & \sin\theta\\
         -\sin\theta & \cos\theta
        \end{bmatrix}
        \begin{bmatrix}
         x(s) \\
         y(s)
        \end{bmatrix}.
  \end{equation}

For a non-rotated tube, i.e. $\theta=0$, when $\sigma$ is between $\sigma_{t}$ and $\sigma_{c}$ the expression for the force reads
 \begin{eqnarray}\label{stscn234}
  F_{l}(L)&&=2\int_{0}^{s_{\min}}\frac{\mathrm{d}\delta x(s)}{\mathrm{d}s}f^{\parallel}_{{\cal A}}(s_{w},0|L)\mathrm{d}s+\nonumber\\
  &&+2\int_{s_{1}}^{(nT/2)-\vartheta(n\ \mathrm{mod}\ 3)s_{1}}\frac{\mathrm{d} x(s)}{\mathrm{d}s}f^{\parallel}_{{\cal A}}(s_{w},0|L)\mathrm{d}s\nonumber\\&&
  +2\vartheta(n\ \mathrm{mod}\ 3)\int_{(nT/2)-s_{\min}}^{nT/2}\frac{\mathrm{d}\delta x(s)}{\mathrm{d}s}f^{\parallel}_{{\cal A}}(s_{w},0|L)\mathrm{d}s,\ \ \ \ \ \ \ \ \
\end{eqnarray}
where $\vartheta$ denotes the Heaviside strep function, with the condition $\vartheta(0)=0$, and $"\text{mod}"$ is the standard modulo operator which, in the concrete case, returns the remainder after division of $n$ by 3, with $n=2,3,4$. In \eref{stscn234} $s_{w}=s_{\min}$, as the latter satisfies the condition $\mathrm{d}y(s)/\mathrm{d}s|_{s=s_{\min}}=0$, $s_{1}$ is solution of the equation $y(s_{1})=y(0)$ and $\delta x(s)=|x(s)-x(s')|$ with the condition $y(s)=y(s')$.

%------------------------------------------------------------------------------------------------------------------------------------
\begin{table*}[t]
\caption{\label{TableParamExp} Values of the best-fit parameters of the model \eref{modelfit1}. The uncertainty of each is given in parenthesis. The reported values are obtained using the build-in procedure \textsc{NonliearModelFit} in \textsc{Mathematica}$^\circledR$ combined with minimization of the Pearson's chi-square $\left(\chi^{2}\right)$ test.}
\centering
\begin{tabular}{@{}c c c c c c c c c c c c c c @{}}
\toprule[1.2pt]
\multicolumn{2}{c}{}                           & \multicolumn{4}{c}{$n=2$} & \multicolumn{4}{c}{$n=3$} & \multicolumn{4}{c}{$n=4$} \\
\cmidrule(r){3-6}\cmidrule(r){7-10}\cmidrule(r){11-14}
\multicolumn{2}{c}{$\Phi_{l}$} & $\Sigma_{a}$   & $\upsilon$  & $\Sigma_{b}$  & $\tau$  & $\Sigma_{a}$   & $\upsilon$  & $\Sigma_{b}$  & $\tau$  & $\Sigma_{a}$   & $\upsilon$  & $\Sigma_{b}$  & $\tau$  \\ \midrule
\multirow{4}{*}{\rotatebox{90}{$L=L_{m}$}}    & $\theta=0$       & 2.57(9)     & 4.10(1)    & 2.74(9)    & 3.95(3)    & 0.43(1)     & 10.65(1)    & 1.61(2)    & 1.76(2)    & 0.30(1)     & 25.71(1)    & 1.47(1)    & 1.44(1)    \\ \cmidrule(l){2-14}
                              & $\theta=\pi/2$   & -     & -    & 2.14(4)    & 2.20(5)    & -     & -    & -    & -    & -     & -    & -    & -    \\ \cmidrule(l){2-14}
                              & $\theta=\pi/3$   & -     & -    &   -        &  -  & -     & -    & 1.31(4)    & 1.75(4)    & -     & -    & -    & -    \\ \cmidrule(l){2-14}
                              & $\theta=\pi/4$   & -     & -    & -          & -    & -     & -    & -    & -    & -     & -    & 1.24(3)    & 1.46(2)    \\ \midrule[1.2pt]
\multirow{4}{*}{\rotatebox{90}{$L=L_{c}$}}    & $\theta=0$       & -     & -    & 1.76(2)    & 2.65(3)    & -     & -    & 1.21(1)    & 1.93(1)    & -     & -    & 0.90(1)    & 1.79(1)    \\ \cmidrule(l){2-14}
                              & $\theta=\pi/2$   & 0.70(1)     & 7.73(1)    & 2.19(2)    & 1.91(2)    & -     & -    & -    & -    & -     & -    & -    & -    \\ \cmidrule(l){2-14}
                              & $\theta=\pi/3$   & -     & -    & -    & -    & 0.36(1)     & 11.76(1)    & 1.17(1)    & 1.92(1)    & -     & -    & -    & -    \\ \cmidrule(l){2-14}
                              & $\theta=\pi/4$   & -     & -    & -    & -    & -     & -    & -    & -    & -     & -    & 0.96(6)    & 1.74(5)    \\ \bottomrule[1.2pt]
\end{tabular}
\end{table*}
%------------------------------------------------------------------------------------------------------------------------------------

Being in the same range of values for $\sigma$, one can still make use of \eref{sbstn234} to calculate the tube-plate force when $\theta=\pi/n$ with $n=2$, but the corresponding expression for $n=3$ reads
\begin{eqnarray}\label{stscn3}
  F_{l}(L)&&=2\int_{T/2}^{s_{2}}\frac{\mathrm{d} x_{\pi/3}(s)}{\mathrm{d}s}f^{\parallel}_{{\cal A}}\left(s_{w},\left.\frac{\pi}{3}\right\vert L\right)\mathrm{d}s+\nonumber\\
  &&+2\int_{2T-s_{\min}}^{2T}\frac{\mathrm{d}\delta x_{\pi/3}(s)}{\mathrm{d}s}f^{\parallel}_{{\cal A}}\left(s_{w},\left.\frac{\pi}{3}\right\vert L\right)\mathrm{d}s,
\end{eqnarray}
where $s_{\min}$ is determined as described in the text above at $\theta=0$, $s_{2}$ is solution of the equation $y_{\pi/3}(s_{2})=y_{\pi/3}(2T)$ and $\delta x_{\pi/3}(s)=|x_{\pi/3}(s)-x_{\pi/3}(s')|$ with the condition $y_{\pi/3}(s)=y_{\pi/3}(s')$.

Last but not least for $n=4$, at $\theta=\pi/n$ and pressure well above $\sigma_{t}$, \eref{sbstn234} can be used to calculate $F_{l}(L)$, but only until $\sigma$ becomes such that a saddle point $[\kappa(s)=0]$ occurs for some $s$ whose value is between $T$ and $3T/2$. Further increase of $\sigma$ will result in sign change of the curvature $\kappa$ in the specified interval, as $\kappa(s_{\max}')>0$ and $\kappa(s_{\min}')<0$. Hence, for such conditions and geometry of the nanotube the force reads
\begin{eqnarray}\label{stscn4}
  F_{l}(L)&&=2\int_{T/2}^{s_{\max}'}\frac{\mathrm{d} x_{\pi/4}(s)}{\mathrm{d}s}f^{\parallel}_{{\cal A}}\left(s_{w},\left.\frac{\pi}{4}\right\vert L\right)\mathrm{d}s+\nonumber\\
  &&+2\int_{s_{\max}'}^{s_{\min}'}\frac{\mathrm{d}\delta x_{\pi/4}(s)}{\mathrm{d}s}f^{\parallel}_{{\cal A}}\left(s_{w},\left.\frac{\pi}{4}\right\vert L\right)\mathrm{d}s+\nonumber\\
  &&+2\int_{s_{3}}^{3T-s_{\max}'-s_{3}}\frac{\mathrm{d} x_{\pi/4}(s)}{\mathrm{d}s}f^{\parallel}_{{\cal A}}\left(s_{w},\left.\frac{\pi}{4}\right\vert L\right)\mathrm{d}s+\nonumber\\
  &&+2\int_{3T-s_{\min}'}^{3T-s_{\max}'}\frac{\mathrm{d}\delta x_{\pi/4}(s)}{\mathrm{d}s}f^{\parallel}_{{\cal A}}\left(s_{w},\left.\frac{\pi}{4}\right\vert L\right)\mathrm{d}s+\nonumber\\
  &&+2\int_{3T-s_{\max}'}^{5T/2}\frac{\mathrm{d} x_{\pi/4}(s)}{\mathrm{d}s}f^{\parallel}_{{\cal A}}\left(s_{w},\left.\frac{\pi}{4}\right\vert L\right)\mathrm{d}s.
\end{eqnarray}
Here $s_{3}$ is such that $y_{\pi/4}(s_{3})=y_{\pi/4}(s_{\max}')$ and $\delta x_{\pi/4}(s)=|x_{\pi/4}(s)-x_{\pi/4}(s')|$ with the condition $y_{\pi/4}(s)=y_{\pi/4}(s')$.

The numerical evaluation of the above expressions for a standard van der Waals interaction is visualized on Fig. \ref{fig:total_forces}, while the observed dependance is commented in details in Sec. \ref{sec:ResAndDisc}.
%
%=====================
\section{Results and discussion}\label{sec:ResAndDisc}
%=====================
%
In interpreting the data illustrated on Fig. \ref{fig:total_forces}, we propose the following approximation
\begin{eqnarray}\label{modelfit1}
  \Phi_{l}(\hat{\sigma},\theta,L)&&=\Phi_{l}(\hat{\sigma}_{b},L)\Theta(\hat{\sigma}_{b}-\hat{\sigma})+\vartheta(\hat{\sigma}-\hat{\sigma}_{b})\nonumber\\
  &&\times
  \begin{cases}
    \left(\Sigma_{a}\hat{\sigma}^{-1}\right)^{\upsilon}- \left(\Sigma_{b}\hat{\sigma}^{-1}\right)^{\tau}, & \text{in case $(I)$}\\
    -\Sigma_{b}\hat{\sigma}^{-\tau}, & \text{in case $(II)$}
  \end{cases}\ \ \ \ \ \
\end{eqnarray}
where $\hat{\sigma}\equiv\sigma/\sigma_{c}$, $\Theta$ s again the Heaviside step function, but with the convention $\Theta(0)=1$, case $(I)$ refers to systems with $L=L_{m}$ and $\theta=0$ at $n=2,3,4$ or if $L=L_{c}$, when $\theta=\theta/2$ and $\theta/3$, while case $(II)$ describes the geometry $\theta=\pi/n,\ n=2,2,4$ for $L=L_{m}$ together with the one where $L=L_{c}$ with $\theta=0$ and $\theta=\pi/4$. The values of the parameters $\Sigma_{a},\ \Sigma_{b},\ \upsilon\ \text{and}\ \tau$ are given in Table \ref{TableParamExp}. The first term in \eref{modelfit1} mirrors the observation that for $\hat{\sigma}\leq\hat{\sigma}_{b}$ the cross section contour does not change, and hence the dependance of $\Phi_{l}$ from $\hat{\sigma}$, at fixed separation $L$, is constant equal to $\Phi_{l}(\hat{\sigma}_{b})=-8.28\times 10^{-3}$ when $L_{m}/\rho=L_{c}/\rho=2$. The second part, case $(I)$, is chosen in analogy with the Mie potential \cite{MG1903}, describing the intermolecular repulsion at short distances and the attraction at large. Here, we observe the occurrence of a minimum in the $\Phi_{l}(\hat{\sigma})$-dependance both for $L=L_{m}$ at $\theta=0$ and any of the considered values of $n$ as well as for $L=L_{c}$ at $\theta=\pi/n$ with $n=2\ \text{and}\ 3$. In understanding the so-constructed "repulsive" and "attractive" terms (proportional to $\Sigma_{a}^{\upsilon}$ and $\Sigma_{b}^{\tau}$, respectively) it is worth describing the change of the CNT geometry upon deformation and link each step of it with the observed behaviour of $\Phi_{l}(\hat{\sigma})$.

As noted in the last paragraph of Sec. \ref{sec:CrossSec}, the lowest value of $\sigma$ above which one can observe non-circular cross section profile of a CNT is $3$, which corresponds to contour with symmetry index $n=2$ [Fig. \ref{fig:total_forces}$\mathbf{(a)}$]. In this case, the increase of $\hat{\sigma}$ above $\hat{\sigma}_{b,n=2}=0.572$ flattens the tube. If the realized geometry corresponds to $\theta=0$ and one fixes $L=L_{m}$, surface elements, both facing "towards" and "away" from the substrate's surface, which for lower values of $\hat{\sigma}$ were distant from the plate, will now appear closer. Since, $\Phi_{l}$ is proportional to $\oint_{S}\Delta_{\theta}^{-\nu}(s){\rm{d}}S$ [refer to expressions Eqs. (\ref{SIAgeneralsimple}) and (\ref{FpuA})], the decrease of the separations $\Delta_{0}(s)$, due to the altered geometry, in comparison to these of a circular cross section, renders force which magnitude is higher than $\Phi_{l}(\hat{\sigma}_{b})$. Our study shows that $\Phi_{l}(\hat{\sigma})$ continues to increase [note the curve {\color{blue_n}{{\Large-}\hspace{-0.1cm}{\Large-}\hspace{-0.1cm}{\small$\blacksquare$}\hspace{-0.09cm}{\Large-}\hspace{-0.1cm}{\Large-}}} on Fig. \ref{fig:total_forces}$\mathbf{(a)}$] even above $\hat{\sigma}_{t,n=2}=0.637$, reaching its maximum at $\hat{\sigma}_{\max,n=2}=0.658$ with a value of $\Phi_{l}(\hat{\sigma}_{{\max},n=2},0,L_{m})=-10.73\times 10^{-3}$. After this point $\Phi_{l}$ decreases gradually towards $\hat{\sigma}_{c}\equiv1$, mainly due to the "indentation" of the tube's middle area elements, which face "towards" the substrate's surface.

On the other hand, if upon "flattening" the tube's cross section profile mimics the one of a CNT with symmetry index $n=2$ rotated to its symmetry angle $\theta=\pi/2$, the non-retarder van der Waals force will be a decreasing function of $\hat{\sigma}$ for any value greater than $\hat{\sigma}_{b,n=2}$ [note the curve {\color{red_n}{{\Large-}\hspace{-0.1cm}{\Large-}\hspace{-0.1cm}{\large$\bullet$}\hspace{-0.09cm}{\Large-}\hspace{-0.1cm}{\Large-}}} on Fig. \ref{fig:total_forces}$\mathbf{(a)}$ together with case $(II)$ of \eref{modelfit1}] with an infinite slope at $\hat{\sigma}=\hat{\sigma}_{b,n=2}$. This behaviour is conditioned by the relatively lower fraction of the tube's surface area which face "towards" the substrate's surface in comparison to that of a non-deformed CNT.

If now, instead of fixed minimal surface-to-surface distance $L_{m}$ between a CNT and a substrate, one takes $L=L_{c}$ [Fig. \ref{fig:total_forces}$\mathbf{(b)}$] the behaviour of the force with respect to the tube's inclination is vice versa to that discussed in the previous paragraphs. The force maximum $\Phi_{l}(\hat{\sigma}_{{\max},n=2},\pi/2,L_{c})=-8.90\times 10^{-3}$ is reached at $\hat{\sigma}_{\max,n=2}=0.60<\hat{\sigma}_{t,n=2}$ when one takes $L_{c}/\rho=2$. This behaviour is easily explained considering that, when the tube's center-to-substrate's surface distance $(L_{c}+\rho)$ is fixed, the separations $\Delta_{0}(s)$ of the CNT's area elements $A_S^{\rm to}$, broaden with the increase of $\hat{\sigma}$ above $\hat{\sigma}_{b,n=2}$. As a result, the force $\Phi_{l}$ weakens monotonically [note the curve {\color{purple_n}{{\Large-}\hspace{-0.1cm}{\Large-}\hspace{-0.1cm}$\blacktriangledown$\hspace{-0.09cm}{\Large-}\hspace{-0.1cm}{\Large-}}} on Fig. \ref{fig:total_forces}$\mathbf{(b)}$]. On the other hand, $\Delta_{\pi/2}(s)$ of a portion of $A_S^{\rm to}$ decreases for a certain interval of values of the pressure, which results in the occurrence of a force maximum [note the curve {\color{orange_n}{{\Large-}\hspace{-0.1cm}{\Large-}\hspace{-0.1cm}$\blacktriangle$\hspace{-0.09cm}{\Large-}\hspace{-0.1cm}{\Large-}}} on Fig. \ref{fig:total_forces}$\mathbf{(b)}$]. In an analogical manner one can describe the force-pressure dependance [Figs. \ref{fig:total_forces}$\mathbf{(c)}-\mathbf{(f)}$]] in a CNT-plate systems with symmetry index $n>2$.

It is worth noting that the curves $\Phi_{l}(\hat{\sigma},0,L)$ and $\Phi_{l}(\hat{\sigma},\pi/n,L)$ tend towards one another when the dissimilarities, between the rotated (at $\theta=\pi/n$) and non-rotated (at $\theta=0$) cross section profiles, decreases with the inflation of $n$. One also observes that $\Phi_{l}(\hat{\sigma}_{{\max},n},\theta,L)$ weakens with the increase of $n$, as for $n=4$ at $L_{c}/\rho=2$ it vanishes [note the curve {\color{orange_n}{{\Large-}\hspace{-0.1cm}{\Large-}\hspace{-0.1cm}$\blacktriangle$\hspace{-0.09cm}{\Large-}\hspace{-0.1cm}{\Large-}}} on Fig. \ref{fig:total_forces}$\mathbf{(f')}$].
%=====================
\section{Summary and concluding remarks}\label{sec:Conclusions}
%=====================
%
The aim of the current article was to study the observed behaviour of the non-retarded van der Waals force between a planar substrate and a SWCNT, when both are immersed in a liquid medium which exerts hydrostatic pressure on the tube's surface, thereby altering its cross section profile. The shape of the SWCNT was described as a continual structure characterized by its symmetry index $n$ - see Sec. \ref{sec:CrossSec}. Here we considered only contours with $n=2,3\ \text{and}\ 4$. Two principle mutual positions of the tube with respect to the substrate were studied:
\begin{itemize}
  \item when one keeps constant the minimal separation $(L_{m})$ between the surfaces of the interacting objects;
  \item when the distance from the CNT's center to the substrates bounding surface $(L_{c}+\rho)$ is fixed.
\end{itemize}
Within these conditions, using the technique of the surface integration approach - see Sec. \ref{sec:SIAapp}, we derived in integral form expressions which give the dependance of the commented force from the applied pressure - see Sec. \ref{sec:SWCNTforce}. The results from the numerical evaluation of these expressions are presented in Fig. \ref{fig:total_forces}, and an explanation for the established dependance is indicated in Sec. \ref{sec:ResAndDisc}.

From what is presented on Fig. \ref{fig:total_forces} it is clear that we choose to vary $\sigma$ only between $\sigma_{b}$ and $\sigma_{c}$, for any realized symmetry of the non-circular contour. This is because presently the description of the CNTs cross section above the "contact" pressure is debatable \cite{MPFAP2012}.

In the last paragraph of Sec. \ref{sec:CrossSec} it was briefly stated that the value of $\sigma_{c}$ is calculated based on the criteria for \textit{overlaping} of the opposite sites of some CNT transverse contour, which in some sense is a purely mathematical consideration. Since nanotubes are discrete atomistic structures, one can estimate the physical value of the "contact" pressure taking into account some upper limit on the minimal separation between opposing atoms on a single contour loop. Thus, if one chooses as a constrain the condition that $\sigma_{c}$ is this value of the hydrostatic pressure at which the carbon-carbon Lennard-Jones potential is zero \cite{note11}, i.e. $V_{\rm LJ}^{\rm C-C}(r_{0})=0$, then we have for a SWCNTs with $n=2$ that: $\sigma_{c}^{(10,10)}=4.145\ (\rho=0.34\ {\rm nm},\ p_{c}\sim19\ {\rm GPa})$, $\sigma_{c}^{(20,20)}=4.620\ (\rho=0.69\ {\rm nm},\ p_{c}\sim3\ {\rm GPa})$, $\sigma_{c}^{(40,40)}=4.910\ (\rho=1.38\ {\rm nm},\ p_{c}\sim0.3\ {\rm GPa})$. Here the appearing superscripts are the values of the chiral indices of the nanotubes. When calculating the so presented values we have assumed that for pair of carbon atoms $r_{0}=3.37\ {\AA}$ \cite{SM2003} and used Eq. (4.3) from Ref. \cite{Vaj2013} to determine the radius $\rho$ of the nanotubes. The corresponding pressure $p_{c}$ in units $\rm GPa$ is also given, assuming for $D$ the value $1.13\ {\rm eV}$ reported in Ref. \cite{Zang2007}.

Last but not least, in scope of arguing the experimental feasibility of the presented theory, we give the magnitude of the force for concrete system of substances, say (40,40) armchair SWCNT and a graphene sheet immersed in water. For such a system we take $A_{\rm Ham}=10\ {\rm zJ}$ \cite{RFCCC2007} and the value for $D$ as considered above. Hence, assuming $n=2$ we have that $p_{b}\sim0.2\ {\rm GPa}$, $p_{t}\sim0.23\ {\rm GPa}$, $p_{c}\sim0.3\ {\rm GPa}$, $F_{l}(p_{b})\simeq-43.5\ {\rm pN}/{\rm \mu m}$ and
\begin{itemize}
  \item for $L=L_{m}$ at $\theta=0$, $p_{\max}\sim0.24\ {\rm GPa}$ and $F_{l}(p_{\max})\simeq-56.3\ {\rm pN}/{\rm \mu m}$;
  \item for $L=L_{c}$ at $\theta=\pi/2$, $p_{\max}\sim0.22\ {\rm GPa}$ and $F_{l}(p_{\max})\simeq-46.7\ {\rm pN}/{\rm \mu m}$.
\end{itemize}

\acknowledgments
%================================================================================
The authors gratefully acknowledge the financial support via Contract No. DN 02/8 of Bulgarian NSF.
%================================================================================
%

\end{document}